\documentclass[sigconf]{acmart}
\AtBeginDocument{%
  \providecommand\BibTeX{{%
    \normalfont B\kern-0.5em{\scshape i\kern-0.25em b}\kern-0.8em\TeX}}}

\usepackage{todonotes}
\usepackage{xcolor}
\usepackage{collcell}
\usepackage{colortbl}
\usepackage[capitalise]{cleveref}
\usepackage{xspace}
\usepackage{fp}
\usepackage{siunitx}
\usepackage{xurl}
\usepackage{tabularx}
\usepackage{subfig}
\usepackage{multirow}
\usepackage{booktabs}
\usepackage{listings}
\usepackage{xcolor}
\usepackage{amsmath}

\definecolor{Gray}{gray}{0.9}

\newcommand{\teachersintraining}{teachers in training\xspace}
\newcommand{\Teachersintraining}{Teachers in training\xspace}
\newcommand{\teacherintraining}{teacher in training\xspace}

\newcommand{\task}{task\xspace}
\newcommand{\tasks}{tasks\xspace}
\newcommand{\Tasks}{Tasks\xspace}
\newcommand{\Task}{Task\xspace}
\newcommand{\tasksOtherTerm}{assignments}

\newcommand{\PassauInGermany}{Passau in Germany\xspace}

\newcommand{\litterbox}{\textsc{LitterBox}\xspace}
\newcommand{\scratch}{\textsc{Scratch}\xspace}
\newcommand{\whisker}{\textsc{Whisker}\xspace}
\newcommand{\drscratch}{\textsc{Dr. Scratch}\xspace}

\newcommand{\rqOne}{\textbf{RQ 1:} How do \teachersintraining create programming \tasks?
}

\newcommand{\rqTwo}{\textbf{RQ 2:} How does tool support affect the \task text and the code?
}

\newcommand{\rqThree}{\textbf{RQ 3:} How do \teachersintraining perceive the tool support?}

\newcommand{\summary}[2]{
        \vspace{2mm}
        \noindent
        \fbox{%
            \parbox{.97\linewidth}{%
                    \textbf{#1 Summary.}
                #2
            }%
        }%
}%

\newcommand{\ctrlquote}[2]{%
\textsf{``\textit{#2}''}~\textsf{\tiny(Ctrl,~P#1)}%
}

\newcommand{\trmtquote}[2]{%
\textsf{``\textit{#2}''}~\textsf{\tiny(Trmt,~P#1)}%
}

\newcommand{\DrawPercentageBar}[1]{%
  \begin{tikzpicture}
    \fill[color=black]   (0.0 , 0.0) rectangle (#1*0.05ex , 1.7ex );
    \fill[color=lightgray] (#1*0.05ex  , 0.0) rectangle (5.0ex, 1.7ex);
  \end{tikzpicture}%
}

\newcommand\printpercent[1]{\FPeval\result{round(#1,1)}\num[round-mode=places,round-precision=1]{\result}~\%}

\newcommand{\numParticipants}{85\xspace}

\newcommand{\numMainStudyParticipants}{59\xspace}
\newcommand{\numMainStudyFemale}{46\xspace}
\newcommand{\numMainStudyMale}{13\xspace}

\newcommand{\numExperiencedUniOrMathsCtrl}{79.3\xspace}
\newcommand{\numExperiencedUniOrMathsTrmt}{80\xspace}
\newcommand{\numExperiencedSchoolCtrl}{62.1\xspace}
\newcommand{\numExperiencedSchoolTrmt}{53.3\xspace}

\newcommand{\numPreStudy}{26\xspace}
\newcommand{\numWorkshopParticipants}{14\xspace}
\newcommand{\numSeminarSTwentyOneParticipants}{12\xspace}
\newcommand{\timeMean}{82.52\xspace}

\newcommand{\percentagePrScTT}{86.4\xspace}
\newcommand{\percentageSeScTT}{6.8\xspace}
\newcommand{\percentageParticipantsWithOtherStudies}{8.5\xspace}
\newcommand{\numCtrl}{29\xspace}

\newcommand{\numTrmt}{30\xspace}

\newcommand{\numAnalysedPrograms}{56\xspace}
\newcommand{\numAnalysedSolutionPrograms}{53\xspace}
\newcommand{\numAnalysedStartPrograms}{3\xspace}
\newcommand{\meanblockcount}{58.42\xspace}
\newcommand{\medianblockcount}{40\xspace}
\newcommand{\meanscriptcount}{7.53\xspace}
\newcommand{\medianscriptcount}{5\xspace}
\newcommand{\meanspritecount}{3.44\xspace}
\newcommand{\medianspritecount}{3\xspace}
\newcommand{\meansumbugpatterns}{0.98\xspace}
\newcommand{\mediansumbugpatterns}{1\xspace}
\newcommand{\meansumsmells}{2.71\xspace}
\newcommand{\mediansumsmells}{2\xspace}
\newcommand{\meansumperfumes}{5.91\xspace}
\newcommand{\mediansumperfumes}{6\xspace}
\newcommand{\percentageProcedureorderstartwithprogramthentasktext}{40.4}
\newcommand{\percentageProcedureorderiterativeapproachortrialanderror}{25}
\newcommand{\percentageProcedureorderstartwithtasktextthenprogram}{15.4}
\newcommand{\percentageProcedurestartingpointbeforetasktextandprogrambrainstormingorideasearch}{50}
\newcommand{\percentageProcedurestartingpointbeforetasktextandprogramtargetsetting}{9.6}
\newcommand{\percentageProcedureexternalinspirationexampletask}{13.5}
\newcommand{\percentageProcedureexternalinspirationseminarmaterial}{13.5}
\newcommand{\percentageProcedureexternalinspirationScratchenvironment}{11.5}

\newcommand{\percentageProcedurechoosespritesandorbackground}{17.3}

\newcommand{\percentageProcedureinsecurity}{7.7}

\newcommand{\percentageProcedureorder}{71.2}
\newcommand{\percentageProcedurestartingpointbeforetasktextandprogram}{57.7}
\newcommand{\percentageProcedureexternalinspiration}{30.8}
\newcommand{\percentagereasonPositivefunctionality}{94.1}
\newcommand{\percentagereasonPositivefunctionalitydetection}{70.6}
\newcommand{\percentagereasonPositivefunctionalityimprovement}{54.9}
\newcommand{\percentagereasonPositivefunctionalityother}{17.6}
\newcommand{\percentagereasonPositiverepresentation}{27.5}

\newcommand{\percentagereasonPositiveadvantagesforstudents}{19.6}

\newcommand{\percentagereasonPositiveotheradvantages}{29.4}
\newcommand{\percentagereasonPositiveotheradvantagestime}{9.8}
\newcommand{\percentagereasonPositiveotheradvantagesadditionalhelpfortaskcreation}{9.8}
\newcommand{\percentagereasonPositiveotheradvantagesaffective}{5.9}
\newcommand{\percentagereasonPositiveotheradvantagesother}{5.9}
\newcommand{\percentagereasonNegativefunctionality}{33.3}
\newcommand{\percentagereasonNegativefunctionalitynosolvingofallproblems}{11.8}
\newcommand{\percentagereasonNegativefunctionalityincorrectanalysis}{7.8}
\newcommand{\percentagereasonNegativefunctionalitynoanalysisofsemantics}{9.8}
\newcommand{\percentagereasonNegativefunctionalitynoautomaticrefactoring}{5.9}
\newcommand{\percentagereasonNegativerepresentation}{33.3}
\newcommand{\percentagereasonNegativerepresentationcomprehensionproblemsregardinghints}{25.5}
\newcommand{\percentagereasonNegativerepresentationusabilityofthetool}{9.8}
\newcommand{\percentagereasonNegativenotchildfriendly}{7.8}
\newcommand{\percentagereasonNegativenone}{27.5}

\newcommand{\generalReflectiondifficultiesprogrammingpValue}{0.038\xspace}
\newcommand{\generalReflectiondifficultiesprogrammingEffectSize}{0.66\xspace}
\newcommand{\meanCtrltaskNumHints}{1.28\xspace}
\newcommand{\medianCtrltaskNumHints}{2\xspace}
\newcommand{\meanTrmttaskNumHints}{1.9\xspace}
\newcommand{\medianTrmttaskNumHints}{2\xspace}
\newcommand{\taskNumHintspValue}{0.01\xspace}
\newcommand{\taskNumHintsEffectSize}{0.33\xspace}

\newcommand{\missingloopsensingpValue}{0.035\xspace}
\newcommand{\missingloopsensingEffectSize}{0.57\xspace}
\newcommand{\meanCtrlsumbugpatterns}{1.26\xspace}
\newcommand{\medianCtrlsumbugpatterns}{1\xspace}
\newcommand{\meanTrmtsumbugpatterns}{0.72\xspace}
\newcommand{\medianTrmtsumbugpatterns}{0\xspace}
\newcommand{\sumbugpatternspValue}{0.034\xspace}
\newcommand{\sumbugpatternsEffectSize}{0.66\xspace}

\newcommand{\bugDensityBlockNumpValue}{0.034\xspace}
\newcommand{\bugDensityBlockNumEffectSize}{0.66\xspace}

\newcommand{\smellDensityBlockNumpValue}{0.022\xspace}
\newcommand{\smellDensityBlockNumEffectSize}{0.68\xspace}

\newcommand{\meanCtrlsumsmells}{3.19\xspace}
\newcommand{\medianCtrlsumsmells}{3\xspace}
\newcommand{\meanTrmtsumsmells}{2.28\xspace}
\newcommand{\medianTrmtsumsmells}{2\xspace}
\newcommand{\sumsmellspValue}{0.022\xspace}
\newcommand{\sumsmellsEffectSize}{0.68\xspace}

\newcommand{\reasonNegativerepresentationcomprehensionproblemsregardinghintspValue}{0.048\xspace}
\newcommand{\reasonNegativerepresentationcomprehensionproblemsregardinghintsEffectSize}{0.62\xspace}

\newcommand{\meannumSubtasks}{2.28\xspace}
\newcommand{\mediannumSubtasks}{2\xspace}
\newcommand{\PercentageOnlyExtendTasktype}{70\xspace}
\newcommand{\percentageanimals}{34.5\xspace}
\newcommand{\percentageballgames}{20.7\xspace}
\newcommand{\percentageeverydaysuchasfood}{17.2\xspace}
\newcommand{\percentagefantasyandspace}{13.8\xspace}

\newcommand{\percentagemusicsandmaths}{3.4\xspace}

\newcommand{\cohensKappa}{0.76\xspace}

\copyrightyear{2023}
\acmYear{2023}
\setcopyright{acmlicensed}\acmConference[ITiCSE 2023]{Proceedings of the 2023 Conference on Innovation and Technology in Computer Science Education V. 1}{July 8--12, 2023}{Turku, Finland}
\acmBooktitle{Proceedings of the 2023 Conference on Innovation and Technology in Computer Science Education V. 1 (ITiCSE 2023), July 8--12, 2023, Turku, Finland}
\acmPrice{15.00}
\acmDOI{10.1145/3587102.3588809}
\acmISBN{979-8-4007-0138-2/23/07}

\begin{document}

\title[Exploring Programming Task Creation]{Exploring Programming Task Creation of \\ 
Primary School Teachers in Training
}

\author{Luisa Greifenstein}
\email{luisa.greifenstein@uni-passau.de}
\orcid{0000-0002-9707-7762}
\affiliation{%
  \institution{University of Passau}
  \streetaddress{Innstraße 33}
  \city{Passau}
  \country{Germany}
  \postcode{94032}
}

\author{Ute Heuer}
\email{ute.heuer@uni-passau.de}
\orcid{0009-0005-1400-4509}
\affiliation{%
  \institution{University of Passau}
  \streetaddress{Innstraße 33}
  \city{Passau}
  \country{Germany}
  \postcode{94032}
}

\author{Gordon Fraser}
\email{gordon.fraser@uni-passau.de}
\orcid{0000-0002-4364-6595}
\affiliation{%
  \institution{University of Passau}
  \streetaddress{Innstraße 33}
  \city{Passau}
  \country{Germany}
  \postcode{94032}
}

\renewcommand{\shortauthors}{Luisa Greifenstein, Ute Heuer, \& Gordon Fraser}

\begin{abstract}
  Introducing computational thinking in primary school curricula
  implies that teachers have to prepare appropriate lesson
  material. Typically this includes creating programming \tasks, which
  may overwhelm primary school teachers with lacking programming
  subject knowledge. Inadequate resulting example code may negatively
  affect learning, and students might adopt bad programming habits or
  misconceptions.
  To avoid this problem, automated program analysis tools have the
  potential to help scaffolding task creation processes. For example,
  static program analysis tools can automatically detect both good and
  bad code patterns, and provide hints on improving the code.
  To explore how teachers generally proceed when creating programming
  tasks, whether tool support can help, and how it is perceived by
  teachers, we performed a pre-study with \numPreStudy and a main
  study with \numMainStudyParticipants \teachersintraining and the
  \litterbox static analysis tool for \scratch. We find that
  \teachersintraining (1) often start with brainstorming thematic
  ideas rather than setting learning objectives, (2) write code before
  the \task text, (3) give more hints in their task texts and create fewer bugs when supported by \litterbox, and (4) mention both positive aspects of the tool and suggestions for improvement. These findings provide an improved understanding of how to inform teacher training with respect to support needed by teachers when creating programming \tasks.
\end{abstract}





\maketitle


\bibliographystyle{ACM-Reference-Format}

\section{Introduction}
\label{sec:intro}

\begin{figure}[t!]
	\centering
	\includegraphics[trim={3.2cm 0 0 0},clip,width=1\columnwidth]{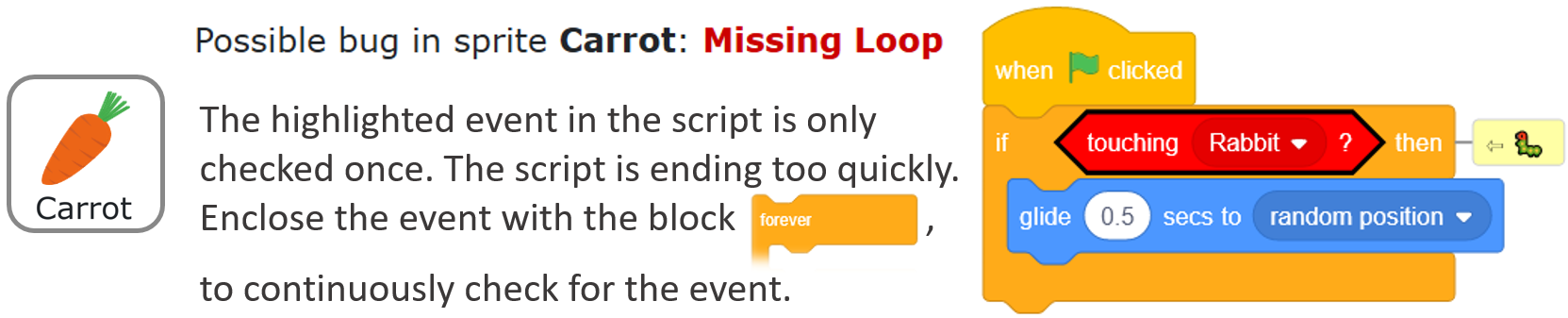}
	\vspace{-2em}
    \caption{\label{fig:examplehint} Hint on the the bug pattern \textit{Missing Loop} of a teacher's \scratch program provided by \litterbox
    .}
	\vspace{-1em}
\end{figure}

Programming \tasks can be used to foster computational thinking~\cite{mannila2014computational} and represent a crucial part of computer science lessons~\cite{ruf2015classification}. Such \tasks usually show example code to young learners before they create their own code, such as in the Use-Modify-Create framework~\cite{lee2011computational}, the PRIMM approach~\cite{sentance2019teachers}, or the TIPP\&SEE strategy~\cite{salac2020tipp}. 
When this example code, however, is of poor quality, learners might adopt bad programming habits and their learning might be impeded
~\cite{hermans2016code}.
This problem is further exacerbated at primary school level:
Computer science related topics have often only recently been introduced to the curricula \cite{heintz2016,nenner2022informatics} and such changes often go along with various issues~\cite{larke2019agentic,wolff2020state,ryder2015being}. Furthermore, particularly primary school teachers consider their lacking subject knowledge a challenge~\cite{sentance2017computing,greifenstein2021challenging}. As a consequence, primary school teachers may struggle when creating example code for their programming \tasks.

Professional programmers receive feedback on their code from automated analysis tools. In principle, such tools may also provide scaffolding for teachers, in particular since new tools have also started to emerge for educational programming languages such as \scratch~\cite{maloney2010scratch}. For example,  \litterbox  provides hints on how to improve code that contains smells or bugs~\cite{fraser2021litterbox,greifenstein2021effects}. \Cref{fig:examplehint} shows a bug in a teacher's example code which would certainly lead to confusion with learners, and the feedback \litterbox provides.

In this paper, we aim to investigate whether and how the use of a tool such as \litterbox also affects the creation of programming \tasks. We therefore conducted an A/B study with \numParticipants teachers in training. All participants were instructed to create a programming \task for primary school children and completed a survey afterwards, and half the participants used the \litterbox tool during the creation. By comparing the resulting \tasks and survey responses we aim to empirically answer the following research questions: 

\smallskip
\noindent \rqOne

\noindent \rqTwo

\noindent \rqThree
\smallskip

We find that (1) teachers tend to start with programming before writing a task text
; (2) tool support has positive effects on code as well as task texts; and (3) teacher feedback helps us provide criteria for effective (tool) support. Consequently, we derive recommendations to inform teacher training on task creation.
\section{Background}

\subsection{Pedagogical Programming \Tasks}

Exploration of how teachers create \tasks should consider criteria for pedagogical \tasks. 
Generally, pedagogical \tasks are characterised by 
the school setting in contrast to real-world or target tasks~\cite{nunan2004task}. 
However, students should be able to transfer their gained competencies which is why pedagogical \tasks should also relate to real-world activities. 
Moreover, pedagogical \tasks should involve the learner and have clear \task specifications~\cite{nunan2004task,goodyear2015teaching}.
Creating such pedagogical \tasks can be seen as one essential part of designing lessons besides organising social structures and designing supportive environments~\cite{goodyear2015teaching}. 
\Tasks also have a crucial role 
in the computer science classroom, e.g., in the form of exercises or examples~\cite{ruf2015classification} which is also reflected in courses for primary school teachers~\cite{geldreich2018off}. 
However, to our knowledge, there has not yet been a systematic evaluation of the creation of pedagogical programming \tasks.


As computational thinking can be fostered by programming activities~\cite{mannila2014computational}, programming \tasks are commonly created by computer science teachers~\cite{ruf2015classification}. 
Independently of their classification (e.g., 
their inherent activities~\cite{bower2008taxonomy}, or the representation of the problem and the solution~\cite{ruf2015classification}), code is often included to be debugged, completed or tested~\cite{bower2008taxonomy,ruf2015classification}. 
This goes along with several teaching approaches that start with or even focus on \tasks with given code such as the Use-Modify-Create framework: In the first two of the three phases learners are tasked to deal with given code~\cite{lee2011computational}. The use phase is further structured within the PRIMM approach, where young learners work intensively with the given code as they predict, run and investigate it~\cite{sentance2019teachers}. The learning strategy TIPP\&SEE provides meta-cognitive scaffolding between the use and modify step and has been designed in particular 
to support diverse learners at primary school level~\cite{salac2020tipp}. 
Other approaches for primary school computer science education such as the Universal Design for Learning also consider splitting tasks into minor parts~\cite{israel2020teaching}.

\subsection{Good Example Code}
The existence of \tasks including example code as part of the learning process~\cite{salac2020tipp,sentance2019teachers,lee2011computational} makes it crucial to consider criteria for good code, such as being dedicated to only one (new) element~\cite{kimura1979reading}, being kept simple~\cite{griffin2012debug}, and adhering to criteria related to names, expressions, or decomposition~\cite{stegeman2014towards}.
%
When learners are exposed to example code, they might adopt good programming habits from code examples if these are well written~\cite{richards2000bugs}. 
However, when the code quality is low, learners might not only imitate this coding style: Hermans and Aivaloglou found that code comprehension decreases when learners aged 12 to 14 years are given programs that contain code smells~\cite{hermans2016code}.
This is problematic since understanding code is a prerequisite to, e.g., deliberate on a bug and fix it during debugging activities~\cite{richards2000bugs}.
Such effects might even increase for primary school children%
~\cite{funke2016primary}.
There exist many suggestions on how to achieve high code quality. Therefore, Stegeman et al. analysed handbooks and interviewed instructors to develop an assessment model~\cite{stegeman2014towards}: They derived nine main criteria, such as names, expressions or decomposition.
Consequently, educators should consider code quality when designing example code, but this might be particularly challenging for primary school teachers who often consider their lacking subject knowledge a problem~\cite{sentance2017computing,greifenstein2021challenging}.
%


\subsection{Tool Support for \scratch Programs}
\label{subsec:toolsforscratch}

Code quality can be analysed automatically by tools~\cite{boe2013,fraser2021litterbox,moreno2015dr,ota2016ninja,techapaloku2017b}, which differ regarding the given feedback~\cite{narciss2013designing}: 
\drscratch~\cite{moreno2015dr} and \litterbox~\cite{fraser2021litterbox}, which are both accessible via a convenient web frontend, both give positive feedback (in terms of scores by \drscratch and code perfumes~\cite{obermuller2021code} by \litterbox) and feedback on concepts. 
While \drscratch returns general information on the detected computational thinking concepts, \litterbox returns specific hints on how to proceed regarding detected smells and bug patterns. 
Therefore, \drscratch might be particularly effective for extending programs according to computational thinking concepts~\cite{troiano2019my} and \litterbox for repairing and improving programs~\cite{greifenstein2021effects}. \litterbox thus might support creating good example code, which is why in this paper we focus on this tool.

\litterbox~\cite{fraser2021litterbox} detects bad (bugs and smells) and good code patterns (perfumes~\cite{obermuller2021code}). When checking programs on the web interface, hints on the detected patterns are given. As shown in \cref{fig:examplehint}, a typical hint consists of an explanation of the underlying concept and (for bad code patterns) a suggested improvement~\cite{fraser2021litterbox}. While the automated hints have been demonstrated to be helpful for debugging foreign code~\cite{greifenstein2021effects}, it has not been investigated yet whether they support teachers in creating programming \tasks.
\section{Method}
\label{sec:method}

\begin{table*}[t]
\centering
\caption{Data used to answer the RQs.}
\vspace{-1em}
	\label{tab:tableDataRQs}
\begin{tabular}{lp{1.5cm}p{1.3cm}p{7cm}p{5.5cm}}
\toprule

RQ & Study & Group & Qualitative Data & Quantitative Data  \\

\midrule

1 
& main study 
& Trmt, Ctrl 
& \textbullet~textual answers on the procedure (``\textit{How did you proceed?}'') and specific issues (``\textit{Where did you have difficulties and what worked well?}'')
& \textbullet~numerical answers on time (``\textit{How much time did you invest in your own \task?}'') \\

2 
& main study
& Trmt, Ctrl 
& \textbullet~\task text (\task title, sub-\tasks, \task type~\cite{ruf2015classification})
& \textbullet~code metrics and code patterns 
\\

3 
& main study, pre-study
& Trmt 
& \textbullet~textual answers on positive tool evaluation (``\textit{Note positive aspects, purposes, advantages, especially helpful hints etc.}'')  and 
negative 
(``\textit{Note negative aspects, limitations, suggestions for improvement, less helpful hints etc.}'')  
& \textbullet~evaluation of the tool's usefulness on a 5-point Likert scale from ``\textit{completely true}'' to ``\textit{not at all true}'' (\textit{``LitterBox can help teachers to create good example programs.''}) \\


\bottomrule
\end{tabular}
\end{table*}

\subsection{Pre-Study}

We performed a preliminary study to optimise the setting of our main study and retrieve additional feedback on the tool.
The pre-study was performed as part of an earlier implementation of the seminar ``Computational Thinking in
Primary School'' at the University of \PassauInGermany with \numSeminarSTwentyOneParticipants \teachersintraining and additionally in a standalone workshop with \numWorkshopParticipants \teachersintraining.
All participants were instructed to use \litterbox for creating a \task and therefore received a short instruction on the usage of \litterbox (but none regarding code quality). In contrast to the main study, participants were completely free in how often to use the tool, they had less time (about 40 instead of 60-70 minutes) and were allowed to work in groups.

The pre-study informed the design of the main study (e.g., regarding time and tool usage) and some updates of the \litterbox tool. The updates enable us to compare the evaluations of earlier versions of the tool in the pre-study with more current versions in the main study in RQ 3 (\cref{tab:tableDataRQs}).

\subsection{Participants}

In the main study, \numMainStudyParticipants teachers in training (\numMainStudyFemale females and \numMainStudyMale males) participated. All of them are teachers in training at the University of \PassauInGermany: \percentagePrScTT~\% are pursuing a degree in primary school education, \percentageSeScTT~\% in secondary school education and \percentageParticipantsWithOtherStudies~\% additionally or exclusively in (media) pedagogy. Within their studies, they chose the seminar ``Computational Thinking in Primary School''.
The \teachersintraining were divided into two groups (Ctrl, Trmt) by their last names. This method was chosen due to the learning platform used. We ensured that the groups are balanced which resulted in \numCtrl participants in group Ctrl and \numTrmt participants in group Trmt and similar programming experience (\cref{tab:experience}). 
While group Ctrl did not receive any additional support, group Trmt were instructed to use the \litterbox tool at least two times. As the \teachersintraining had not used the tool previously, group Trmt also received a short video about its usage and how it works, as we knew from our pre-study and a related study with high school students~\cite{marwan2021promoting} that explaining these aspects would avoid confusion. 

\begin{table}[t]
\centering
\caption{Participants' prior experience with programming.}
\vspace{-1em}
	\label{tab:experience}
\begin{tabular}{lrrr}
\toprule

Group & School & University & Seminar Units 1--4 \\

\midrule

Ctrl & \numExperiencedSchoolCtrl \% & \numExperiencedUniOrMathsCtrl \% & 100 \% \\

Trmt & \numExperiencedSchoolTrmt \% & \numExperiencedUniOrMathsTrmt.0 
\% & 100 \% \\

\bottomrule
\end{tabular}
\end{table}

\subsection{Data Collection}
\Cref{tab:tableDataRQs} gives an overview of the collected data regarding each research question.
All data were collected in the fifth unit of the seminar, in which the theory of scaffolding, the Use-Modify-Create framework~\cite{lee2011computational} and different \task types~\cite{ruf2015classification,geldreich2019aufgabe} were also introduced.
The \teachersintraining were then tasked to create their own \scratch programming \task (consisting of a task text, a starter program, and where applicable a sample solution) for the target group of primary school children. While there were no specifications regarding, e.g., program size or \task type, they were instructed to spend at least 60 to 70 minutes to create their \task. The \teachersintraining were also advised to use a template for their \task text and to design an appealing \task. They were not instructed to use specific programming concepts, but were reminded to bear their self-selected learning objectives such as a specific pattern or control structure in mind. 
%
After completing their \task, they submitted the \task text and a starter \scratch program. When the chosen \task type required changing or extending the code, a solution \scratch program was also submitted.
Finally, they answered a survey consisting of four open and two closed questions for group Trmt respectively two open and one closed question for group Ctrl~(\cref{tab:tableDataRQs}).
The anonymised usage of the collected data has been granted by all teachers in training.

\subsection{Data Analysis}

\subsubsection{Qualitative analysis}
We applied thematic analysis~\cite{bergman2010hermeneutic} on the answers to the open questions of the survey (\cref{tab:tableDataRQs}). One author and one assistant read all statements and agreed on a coding scheme. Then one author rated all data and the assistant rated 20~\% of all data independently to guarantee inter-rater reliability ($K$ = \cohensKappa).
To recognise more generalisable tendencies for our target group of primary school teachers (in training), we counted occurrences for each (sub-)category, which provides further quantitative data. 

\subsubsection{Quantitative analysis}
We used a Wilcoxon rank sum test to measure statistical differences between groups Ctrl and Trmt with $\alpha = 0.05$, and the Vargha-Delaney $\hat{A}_{12}$ effect size, which ranges from 0 to 1. If $\hat{A}_{12} = .50$, there are no effects in favour of any group, if $\hat{A}_{12} > .50$ the values of the dependent variable are higher for group Ctrl than group Trmt and vice versa if $\hat{A}_{12} < .50$.

\subsubsection{RQ 1}
We analysed the answers on the open questions for RQ~1 (\cref{tab:tableDataRQs}) qualitatively and then quantitatively to discover differences between group Ctrl and group Trmt.

\subsubsection{RQ 2}
We analysed the data for RQ 2 (\cref{tab:tableDataRQs}) quantitatively to discover differences between group Ctrl and group Trmt. 
The created \tasks provided us with metrics on the code (calculated with the \litterbox version from 5 January 2023) and the number of types of programming \tasks (deduced from \cite{ruf2015classification}).
As the \teachersintraining were free to choose their \task type, we analysed either the solution \scratch program or the starter \scratch program depending on the chosen \task type: For USE \tasks, there is no solution program and for MODIFY \tasks, the starter programs might contain intended bugs or incomplete code patterns on purpose which would distort the analysis. This resulted in \numAnalysedPrograms analysed programs (\numAnalysedSolutionPrograms solution programs and \numAnalysedStartPrograms starter programs).

\subsubsection{RQ 3}
We analysed the answers on the open questions for RQ~3 (\cref{tab:tableDataRQs}) qualitatively and then quantitatively to discover differences between the pre- and the main study.

\subsection{Threats to Validity}

\subsubsection{External validity}
The teachers in training chose the seminar voluntarily for their studies. This implies that the cohort (1) is self-selected and (2) results might be different for less interested teachers in training on the one hand and more experienced teachers or also larger programs on the other.
%
While participants were instructed to create their own \scratch programs, they were told to use a template for their \task text that also contained an example \task. The example \task was an extend \task type with two subtasks and one hint for each subtask and was aimed at collision detection. While the example was perceived as helpful, it probably led to \teachersintraining creating similar \tasks, which is why the resulting \task texts may not be generalisable to other contexts.


\subsubsection{Internal validity}
The study started with a short explanation that students can learn from the example code in the \tasks
which is why the code should be correct, readable and not promote misconceptions. However, this probably increased the attention to code quality 
 thus subsuming some of the impact of the tool support.
%
While group Trmt were instructed to use the \litterbox tool at least twice, we did not check their behaviour. Moreover, they were not obliged to implement the suggestions of \litterbox. A more formal setting or including how much they actually used the output of a tool could mitigate this threat.

\section{Results}

To answer the research questions posed in \cref{sec:intro}, we
consider the \task texts, \scratch programs and survey results as described in
\cref{sec:method}.

\subsection{RQ 1: Procedure}
\begin{table}[t]\centering
	\caption{Procedure of \teachersintraining to create a task. 
 }
	\vspace{-1em}
	\label{tab:procedure}
	\resizebox{\columnwidth}{!}{
\begin{tabular}{lrl}
    \toprule
	(Sub-)Category & \multicolumn{2}{l}{\% \teachersintraining} \\
 
	\midrule

     \rowcolor{Gray} Starting point & \printpercent{\percentageProcedurestartingpointbeforetasktextandprogram} & \DrawPercentageBar{\percentageProcedurestartingpointbeforetasktextandprogram} \\
    \hspace{0.5em} brainstorming or idea search & \printpercent{\percentageProcedurestartingpointbeforetasktextandprogrambrainstormingorideasearch} & \DrawPercentageBar{\percentageProcedurestartingpointbeforetasktextandprogrambrainstormingorideasearch} \\
    \hspace{0.5em} setting of learning objectives & \printpercent{\percentageProcedurestartingpointbeforetasktextandprogramtargetsetting} & \DrawPercentageBar{\percentageProcedurestartingpointbeforetasktextandprogramtargetsetting} \\

    \rowcolor{Gray} Inspiration & \printpercent{\percentageProcedureexternalinspiration} & \DrawPercentageBar{\percentageProcedureexternalinspiration} \\
    \hspace{0.5em} example \task & \printpercent{\percentageProcedureexternalinspirationexampletask} & \DrawPercentageBar{\percentageProcedureexternalinspirationexampletask} \\
    \hspace{0.5em} seminar material & \printpercent{\percentageProcedureexternalinspirationseminarmaterial} & \DrawPercentageBar{\percentageProcedureexternalinspirationseminarmaterial} \\    
    \hspace{0.5em} \scratch environment & \printpercent{\percentageProcedureexternalinspirationScratchenvironment} & \DrawPercentageBar{\percentageProcedureexternalinspirationScratchenvironment} \\
    
    \rowcolor{Gray} Sprite selection & \printpercent{\percentageProcedurechoosespritesandorbackground} & \DrawPercentageBar{\percentageProcedurechoosespritesandorbackground} \\

    \rowcolor{Gray} Order & \printpercent{\percentageProcedureorder} & \DrawPercentageBar{\percentageProcedureorder} \\
    \hspace{0.5em} start with program, then \task text & \printpercent{\percentageProcedureorderstartwithprogramthentasktext} & \DrawPercentageBar{\percentageProcedureorderstartwithprogramthentasktext} \\
    \hspace{0.5em} iterative approach or trial and error & \printpercent{\percentageProcedureorderiterativeapproachortrialanderror} & \DrawPercentageBar{\percentageProcedureorderiterativeapproachortrialanderror} \\
    \hspace{0.5em} start with \task text, then program & \printpercent{\percentageProcedureorderstartwithtasktextthenprogram} & \DrawPercentageBar{\percentageProcedureorderstartwithtasktextthenprogram} \\

   	\rowcolor{Gray} Insecurity or difficulties & \printpercent{\percentageProcedureinsecurity} & \DrawPercentageBar{\percentageProcedureinsecurity} \\


\bottomrule
\end{tabular}
}
\end{table}

On average, the \teachersintraining spent \timeMean minutes to create their \task.
\cref{tab:procedure} shows the (sub-)categories identified in the responses about how they proceeded when creating their \tasks.

\subsubsection{Starting point}

Before starting the actual task creation, \teachersintraining tend to brainstorm ideas (\cref{tab:procedure}).
The importance of setting learning objectives, however, seems to be underestimated by \teachersintraining (\cref{tab:procedure}). 
Maybe it is (1) considered an obvious part of creating a \task that it does not have to be mentioned; (2) done very quickly and considered a minor and thus not noteworthy part of creating a \task; or (3) forgotten about and thus not considered in the answer. 
The latter two explanations would be problematic as setting goals and related criteria is a crucial aspect when carefully designing tasks~\cite{fuller2007developing}, for example for assessing learning outcomes.

\subsubsection{Inspiration}
The \teachersintraining often required some kind of inspiration when brainstorming ideas (\cref{tab:procedure}): \trmtquote{66}{I looked at all examples again that we have encountered so far. It was not easy for me to come up with my own idea.}
Besides the example \task or other seminar material, some \teachersintraining took inspiration from \scratch sprites, backgrounds, and tutorials.

\subsubsection{Sprite selection}
Sprites are not only a source of inspiration, but their selection can be seen as an explicit step of task creation~(\cref{tab:procedure}), which shows the non-negligible role of figures in \scratch.

\subsubsection{Order}
There are two alternatives of where to start creating the task content: with programming, or with writing the \task text. Most \teachersintraining start---and partially also end---with the code, for example one participant \ctrlquote{46}{created the initial scenario in \scratch; wrote the task description; created the solution}.
%
Additionally, trial and error, iterative, and even parallel approaches are partially used (\cref{tab:procedure}) like \trmtquote{16}{at the same time, I thought about possible \task texts and noted hints for the solution}.

\subsubsection{Insecurity}
Some \teachersintraining perceived the creation of \tasks as difficult (\cref{tab:procedure}), in particular because it was their first attempt to design a complete \task on their own. This matches the finding that especially primary school teachers consider their lacking knowledge a challenge~\cite{sentance2017computing}.

\subsubsection{Specific issues}
While we found no significant differences in the previous categories, when asked if they had specific issues, group Trmt reported significantly fewer programming difficulties ($p = \generalReflectiondifficultiesprogrammingpValue$, $\hat{A}_{12} = \generalReflectiondifficultiesprogrammingEffectSize$) and group Ctrl stated that \ctrlquote{32}{creating compact code} and \ctrlquote{33}{programming efficiently} was difficult. 

\summary{RQ 1}{Before creating their own tasks, the \teachersintraining often search for ideas and take inspiration from other examples. Generally, \teachersintraining tend to start with programming and write the \task text afterwards.}
\subsection{RQ 2: Effects of Tool Support on the \Tasks}

All \tasks (both task texts and \scratch programs) of both groups are available at \url{https://doi.org/10.6084/m9.figshare.22657402}.

\subsubsection{\Task text}

The example task consisted of two subtasks (both were `extend' task types) and two hints. This should be kept in mind when interpreting the following results.
The \tasks of the \teachersintraining had \meannumSubtasks subtasks on average with a median of \mediannumSubtasks. 
Of the \teachersintraining, \PercentageOnlyExtendTasktype~\%  created only `extend' \tasks in all of their subtasks. Other \task types~\cite{ruf2015classification} that were implemented in subtasks could be classified as `create with given code', `debug' and `test' \task types. `Optimising' or `explaining', e.g., were never realised even though these \task types were presented in the seminar.
Regarding the \task titles, the topics seem to focus on animals (\percentageanimals~\%), ball games (\percentageballgames~\%), every day life such as food (\percentageeverydaysuchasfood~\%) and fantasy worlds (\percentagefantasyandspace~\%). However, other school subjects than sports were only addressed in \percentagemusicsandmaths~\% of \task titles.
Group Trmt gave significantly more hints in their \task text than group Ctrl ($p = \taskNumHintspValue$, $\hat{A}_{12} = \taskNumHintsEffectSize$). Group Ctrl inserted \meanCtrltaskNumHints ($\tilde x = \medianCtrltaskNumHints$)  and group Trmt \meanTrmttaskNumHints ($\tilde x = \medianTrmttaskNumHints$) hints on average. This might relate to group Trmt experiencing the tool's hints as helpful and discovering strategies for elaborated feedback on concepts and on how to proceed. They might therefore also want to scaffold their own \tasks more.

\subsubsection{Code metrics and code patterns}

The created \scratch programs contain \meanspritecount sprites ($\tilde x = \medianspritecount$), \meanscriptcount scripts ($\tilde x = \medianscriptcount$) and \meanblockcount blocks ($\tilde x = \medianblockcount$) on average. 
Considering the patterns found in the code, the programs contain on average \meansumbugpatterns different bug patterns ($\tilde x = \mediansumbugpatterns$), \meansumsmells smells ($\tilde x = \mediansumsmells$) and \meansumperfumes perfumes ($\tilde x = \mediansumperfumes$). 
However, the number of smells and bug patterns differs significantly between group Ctrl and group Trmt: When supported by \litterbox, \teachersintraining inserted significantly fewer bug patterns ($p = \sumbugpatternspValue$, $\hat{A}_{12} = \sumbugpatternsEffectSize$) and smells ($p = \sumsmellspValue$, $\hat{A}_{12} = \sumsmellsEffectSize$). Group Ctrl inserted \meanCtrlsumbugpatterns ($\tilde x = \medianCtrlsumbugpatterns$) bug patterns while Group Trmt inserted \meanTrmtsumbugpatterns ($\tilde x = \medianTrmtsumbugpatterns$) bug patterns. Group Ctrl inserted \meanCtrlsumsmells ($\tilde x = \medianCtrlsumsmells$) smells while group Trmt inserted \meanTrmtsumsmells ($\tilde x = \medianTrmtsumsmells$) smells. 
This suggests that the \litterbox tool is helpful for debugging programs and for improving the code quality. To avoid that this finding is only an artefact of smaller programs, we also calculated the bug density and smell density with $\#$\textit{bugs}$/\#$\textit{blocks} and $\#$\textit{smells}$/\#$\textit{blocks} which is still significantly lower for group Trmt than group Ctrl (bugs: $p = \bugDensityBlockNumpValue$, $\hat{A}_{12} = \bugDensityBlockNumEffectSize$; smells: $p = \smellDensityBlockNumpValue$, $\hat{A}_{12} = \smellDensityBlockNumEffectSize$). 
We also looked at the individual patterns and (despite the small sample size) found that the \textit{Missing Loop} bug pattern occurs significantly more often in the programs of group Ctrl ($p = \missingloopsensingpValue$, $\hat{A}_{12} = \missingloopsensingEffectSize$). This could be because the bug pattern is generally very common and the corresponding \litterbox hint (\cref{fig:examplehint}) is easy to implement.

\summary{RQ 2}{\litterbox supports \teachersintraining with creating more scaffolded \task texts and less faulty programs while there were no effects on, e.g., program size or the \task type.}
\subsection{RQ 3: Tool Evaluation}

\begin{table}[t]\centering
	\caption{Evaluation of positive aspects of the \litterbox tool for creating \tasks
    . 
 }
	\vspace{-1em}
	\label{tab:toolpositiveevaluation}
\begin{tabular}{lrllllll}
    \toprule
	(Sub-)Category & \multicolumn{2}{l}{\% \teachersintraining} \\
 
	\midrule

    \rowcolor{Gray} Functionality & \percentagereasonPositivefunctionality~\% & \DrawPercentageBar{\percentagereasonPositivefunctionality} \\
    \hspace{0.5em} detection of patterns & \percentagereasonPositivefunctionalitydetection~\% & \DrawPercentageBar{\percentagereasonPositivefunctionalitydetection} \\
    \hspace{0.5em} suggestions for improvement & \percentagereasonPositivefunctionalityimprovement~\% & \DrawPercentageBar{\percentagereasonPositivefunctionalityimprovement} \\
    \hspace{0.5em} other & \percentagereasonPositivefunctionalityother~\% & \DrawPercentageBar{\percentagereasonPositivefunctionalityother} \\

    \rowcolor{Gray} Advantages for teachers & \percentagereasonPositiveotheradvantages~\% & \DrawPercentageBar{\percentagereasonPositiveotheradvantages} \\
    \hspace{0.5em} time & \percentagereasonPositiveotheradvantagestime~\% & \DrawPercentageBar{\percentagereasonPositiveotheradvantagestime} \\
    \hspace{0.5em} additional help & \percentagereasonPositiveotheradvantagesadditionalhelpfortaskcreation~\% & \DrawPercentageBar{\percentagereasonPositiveotheradvantagesadditionalhelpfortaskcreation} \\
    \hspace{0.5em} affective help & \percentagereasonPositiveotheradvantagesaffective~\% & \DrawPercentageBar{\percentagereasonPositiveotheradvantagesaffective} \\
    \hspace{0.5em} other & \percentagereasonPositiveotheradvantagesother~\% & \DrawPercentageBar{\percentagereasonPositiveotheradvantagesother} \\

    \rowcolor{Gray} Representation & \percentagereasonPositiverepresentation~\% & \DrawPercentageBar{\percentagereasonPositiverepresentation} \\

    \rowcolor{Gray} Usefulness for students & \percentagereasonPositiveadvantagesforstudents~\% & \DrawPercentageBar{\percentagereasonPositiveadvantagesforstudents} \\

\bottomrule
\end{tabular}
\end{table}
\begin{table}[t]\centering
	\caption{Evaluation of negative aspects of the \litterbox tool for creating \tasks
    .
}
	\vspace{-1em}
	\label{tab:toolnegativeevaluation}
\begin{tabular}{lrllllll}
    \toprule
	(Sub-)Category & \multicolumn{2}{l}{\% \teachersintraining} \\
 
	\midrule

    \rowcolor{Gray} Functionality & \percentagereasonNegativefunctionality~\% & \DrawPercentageBar{\percentagereasonNegativefunctionality} \\
    \hspace{0.5em} no solving of all problems & \percentagereasonNegativefunctionalitynosolvingofallproblems~\% & \DrawPercentageBar{\percentagereasonNegativefunctionalitynosolvingofallproblems} \\ 
    \hspace{0.5em} no analysis of semantics & \percentagereasonNegativefunctionalitynoanalysisofsemantics~\% & \DrawPercentageBar{\percentagereasonNegativefunctionalitynoanalysisofsemantics} \\
    \hspace{0.5em} incorrect analysis & \percentagereasonNegativefunctionalityincorrectanalysis~\% & \DrawPercentageBar{\percentagereasonNegativefunctionalityincorrectanalysis} \\
    \hspace{0.5em} no automatic refactoring & \percentagereasonNegativefunctionalitynoautomaticrefactoring~\% & \DrawPercentageBar{\percentagereasonNegativefunctionalitynoautomaticrefactoring} \\
    
    \rowcolor{Gray} Representation & \percentagereasonNegativerepresentation~\% & \DrawPercentageBar{\percentagereasonNegativerepresentation} \\
    \hspace{0.5em} comprehension problems & \percentagereasonNegativerepresentationcomprehensionproblemsregardinghints~\% & \DrawPercentageBar{\percentagereasonNegativerepresentationcomprehensionproblemsregardinghints} \\
    \hspace{0.5em} usability & \percentagereasonNegativerepresentationusabilityofthetool~\% & \DrawPercentageBar{\percentagereasonNegativerepresentationusabilityofthetool} \\

    \rowcolor{Gray} None & \percentagereasonNegativenone~\% & \DrawPercentageBar{\percentagereasonNegativenone} \\

    \rowcolor{Gray} No usefulness for students & \percentagereasonNegativenotchildfriendly~\% & \DrawPercentageBar{\percentagereasonNegativenotchildfriendly} \\

\bottomrule
\end{tabular}
\end{table}
\begin{figure}[tb]
	\centering
	\includegraphics[width=\columnwidth]{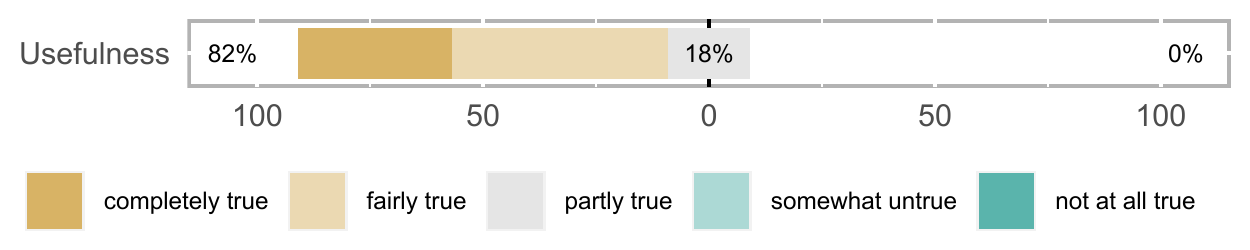}
	\caption{\label{fig:boxplotUsefulness}Rated usefulness of the tool for creating \tasks.}
\end{figure}

The categories of the textual answers provide information about positive (\cref{tab:toolpositiveevaluation}) and negative aspects (\cref{tab:toolnegativeevaluation}) of the tool support.

\subsubsection{General evaluation}
Overall, the \teachersintraining were in favour of the tool support and no \teacherintraining disagreed with its usefulness (\cref{fig:boxplotUsefulness}).
This aligns with primary school teachers considering tools useful for giving feedback (even though they often were not aware of tools)~\cite{greifenstein2021challenging}.
When directly asked about negative aspects of and suggested improvements for the tool, \percentagereasonNegativenone~\% of \teachersintraining stated that they cannot think of any (\cref{tab:toolnegativeevaluation}).

\subsubsection{Functionality}

\Teachersintraining like that the \litterbox tool detects different types of patterns and gives suggestions for improvement (\cref{tab:toolpositiveevaluation}).
Indeed, most \litterbox hints are composed of feedback on concepts and feedback on how to proceed~\cite{fraser2021litterbox,narciss2013designing}, \trmtquote{82}{that makes it easier for inexperienced teachers to recognise and fix bugs}.
%
\Teachersintraining perceived some missing functionality as negative (\cref{tab:toolnegativeevaluation}). While incorrect analyses such as false positives are a common problem of automated feedback tools and should therefore be explained beforehand to enable users to interpret automated hints~\cite{marwan2021promoting}, other criticised aspects are out of the scope of the \litterbox tool such as the analysis of semantics (\cref{tab:toolnegativeevaluation}):  
\trmtquote{35}{Litterbox cannot judge, if my code makes sense semantically}. This is true as \litterbox is a static analysis tool that analyses the code patterns but not the output~\cite{fraser2021litterbox}. 
Semantics could be analysed with dynamic analysis tools such as the \whisker test framework for \scratch~\cite{stahlbauer2019testing}. Another interesting suggestion of the \teachersintraining was to perform automatic refactoring (as for example already explored for \scratch programs~\cite{adler2021improving,techapalokul2019code}) as \trmtquote{61}{you still have to repair the bug yourself and find a new solution}. 
Automatic refactoring might save time for teachers, although one needs to be careful regarding learning processes: Elaborated feedback involves and motivates the learner (both at teacher or student level) more than a provided or automatically generated solution~\cite{narciss2013designing,greifenstein2022common}.

\subsubsection{Advantages for teachers}

\Teachersintraining perceive several advantages especially for teachers such as the additional help they receive, or help in terms of affective factors (\cref{tab:toolpositiveevaluation}). \litterbox might moreover save time as \trmtquote{69}{bugs can be detected without having to search and try for a long time} and you \trmtquote{42}{get feedback in the process and are able to react to it immediately}. 
This refers to some keys of effective feedback such as being timely, actionable and ongoing or formative~\cite{wiggins2012seven}. 
This could support teachers regarding the frequently identified lack of time for planning and during computer science lessons~\cite{sentance2017computing,yadav2016expanding}, as for example reported in a recent study on the efficiency of automated hints for debugging \scratch programs~\cite{greifenstein2021effects}. Apart from timing issues, automated tools might also provide some support regarding the reported relatively low computer science self-esteem~\cite{vivian2020international} and programming self-concept~\cite{greifenstein2021challenging} of primary school teachers as \trmtquote{64}{it is not only shown what is wrong or messy, but also what you have done well. That is motivating} (i.e., it reports code perfumes~\cite{obermuller2021code}).

\subsubsection{Representation}

The way \litterbox represents its findings was mentioned both positively and negatively (\cref{tab:toolpositiveevaluation,tab:toolnegativeevaluation}). The \teachersintraining are in favour of the tool's illustration such as \trmtquote{5}{the visual representation is always great}. 
However, \teachersintraining often had comprehension problems. These are significantly lower in
the main compared to the pre-study ($p = \reasonNegativerepresentationcomprehensionproblemsregardinghintspValue$, $\hat{A}_{12} = \reasonNegativerepresentationcomprehensionproblemsregardinghintsEffectSize$). Further investigation is needed on whether this results from the revised study design or improvements of the tool.

\subsubsection{(Not) for students}

While the \teachersintraining were asked about positive and negative aspects of the \litterbox tool for creating tasks, some \teachersintraining also noted their opinion on the usefulness of \litterbox for students. While some \teachersintraining consider the tool \trmtquote{1}{functional for primary school students to check themselves} and to get a \trmtquote{6}{sense of achievement without the teacher's help}, others state that \trmtquote{67}{for a primary school child, Litterbox would probably seem overwhelming, so it is only useful for the teacher}.
Therefore, more research is needed on whether \litterbox is usable by young learners, how they can be activated, for example using self-explanation prompts~\cite{marwan2019evaluation}, and if more straightforward hints are needed.

\summary{RQ 3}{The \teachersintraining consider it helpful to get feedback on code patterns during the creation of a task but suggest some extensions such as automatic refactoring.}
\section{Discussion}

\subsection{Teacher Training in General}
One way for research to actually propagate to the classroom is to inform teacher training, which is a main strategy to counteract challenges such as lacking knowledge or confidence~\cite{greifenstein2021challenging}.
Mason and Rich~\cite{mason2019preparing} performed a systematic literature review of 21 studies on computing education training for primary school teachers and found that teacher training can be effective both for increasing knowledge and for improving attitudes.
The knowledge teachers gain during training can be categorised into technological knowledge (TK), pedagogical knowledge (PK) and content knowledge (CK) according to the TPACK framework~\cite{herring2016handbook}.
We consider these three categories regarding our results and related work on the support needed by teachers when creating programming \tasks.

\subsection{Pedagogical Knowledge of \Tasks}

We found that teachers in training need ideas for their \tasks (RQ 1). Experienced teachers draw inspiration from existing material~\cite{greifenstein2021challenging} from, e.g., Code.org (\url{https://code.org/}), Teach Computing (\url{https://teachcomputing.org/primary-teachers}), the Raspberry Pi Foundation (\url{https://raspberrypi.org/}), or publicly shared \scratch projects. When selecting existing or creating new programs, gender differences should also be considered, i.e., \trmtquote{42}{I thought of a topic that is appealing for both boys and girls}, which is
 in-line with prior research: While both girls and boys like programs with animals, girls' programs rather involve, e.g., music and dancing and boys' programs, e.g., soccer~\cite{grassl2021data}. 
In our study, shared preferences and those of boys were applied, but girls' preferences are rather underrepresented.
These preferences show a need for differentiation, but also allow for cross-curricular teaching with, e.g., artistic subjects, which again opens up a wide range of ideas for \tasks.

\subsection{Content Knowledge of Programming}
We found that teachers in training often have difficulties when programming (RQs 1 and 2). Teacher training is known to be effective for promoting subject knowledge%
~\cite{mason2019preparing}, for example by letting primary school teachers try out a variety of \task types~\cite{geldreich2018off}. This might not only help with pedagogical knowledge but also with content knowledge, as the teachers are exposed to (appropriate) example code. This could also reduce affective issues, as low confidence is related to lacking subject knowledge~\cite{greifenstein2021challenging}.

\subsection{Technological Knowledge of Tools}

We found that even when primary school teachers in training struggle with programming they can be supported with tools such as \litterbox (RQs 1 to 3). Teachers also have difficulties supporting students during programming~\cite{michaeli2019improving, yadav2016expanding}, which again could be supported with automated analysis tools. However, even in-service teachers are often unaware of available tools, although they consider them useful once they are presented to them~\cite{greifenstein2021challenging}.
We therefore suggest that teacher training should introduce automated analysis tools.
Since programming is the most common approach to foster computational thinking~\cite{ausiku2021preparing}, and \scratch is a very popular programming environment, tools for \scratch could often be meaningfully integrated into teacher training.
\section{Conclusions}

Example code is an essential component of educational programming
tasks, but inadequate example code may negatively affect
learning. Unfortunately, primary school teachers often have
insufficient programming knowledge to produce good example code when
preparing tasks. In this paper we studied whether and how the support
of a static code analysis tool, \litterbox, influences teachers in
training when creating tasks. While the study led to important
suggestions on how to improve \litterbox further, we also found
positive effects on code as well as the task text.

While a primary aim of our study is to inform teacher training, it
also has implications on future research in code analysis tools. For
example, analysis tools are built to simply report all code issues
they encounter. However, depending on the task type there can be
different types of code examples, such as starter code, intentionally
buggy code, or solution code, and each type of program may merit
different types of analyses. Future research could also consider
further tools and types of analysis, as well as the use of deep
learning techniques to analyse not only the code in isolation, but in
conjunction with the corresponding task text.

\begin{acks}
This work is supported by the Federal Ministry of Education and Research
through project “primary::programming” (01JA2021) as part
of the “Qualitätsoffensive Lehrerbildung”, a joint initiative of the
Federal Government and the Länder. The authors are responsible
for the content of this publication.
\end{acks}

\bibliography{literature}

\end{document}